\documentclass[eletronic]{vgtc}                      % final (conference style)
\ifpdf%                                % if we use pdflatex
  \pdfoutput=1\relax                   % create PDFs from pdfLaTeX
  \usepackage{graphicx}                % allow us to embed graphics files
  \DeclareGraphicsExtensions{.pdf,.png,.jpg,.jpeg} % for pdflatex we expect .pdf, .png, or .jpg files
\else%                                 % else we use pure latex
  \usepackage{graphicx}                % allow us to embed graphics files
  \DeclareGraphicsExtensions{.eps}     % for pure latex we expect eps files
\fi%

\graphicspath{{figures/}{pictures/}{images/}{./}} % where to search for the images

\usepackage{microtype}                 % use micro-typography (slightly more compact, better to read)
\PassOptionsToPackage{warn}{textcomp}  % to address font issues with \textrightarrow
\usepackage{textcomp}                  % use better special symbols
\usepackage{mathptmx}                  % use matching math font
\usepackage{times}                     % we use Times as the main font
\usepackage{cite}                      % needed to automatically sort the references
\usepackage{tabu}                      % only used for the table example
\usepackage{booktabs}                  % only used for the table example
\onlineid{0}

\vgtccategory{Research}

\vgtcinsertpkg

\title{Social Virtual Reality: Ethical Considerations and Future Directions for An Emerging Research Space}
\author{Divine Maloney\thanks{e-mail: divinem@clemson.edu}\\
        \scriptsize Clemson University
\and Guo Freeman\thanks{e-mail:guof@clemson.edu}\\ %
     \scriptsize Clemson University %
\and Andrew Robb\thanks{e-mail: arobb@clemson.edu}\\ %
     {\scriptsize Clemson University}}

\abstract{
} % end of abstract

\CCScatlist{
  \CCScatTwelve{Human-centered computing}{Virtual Reality}
}

\begin{document}



\abstract{
The boom of commercial social virtual reality (VR) platforms in recent years has signaled the growth and wide-spread adoption of consumer VR. Social VR platforms draw aspects from traditional 2D virtual worlds where users engage in various immersive experiences, interactive activities, and choices in avatar-based representation. However, social VR also demonstrates specific nuances that extend traditional 2D virtual worlds and other online social spaces, such as full/partial body tracked avatars, experiencing mundane everyday activities in a new way (e.g., sleeping), and an immersive means to explore new and complex identities. The growing popularity has signaled interest and investment from top technology companies who each have their own social VR platforms. Thus far, social VR has become an emerging research space, mainly focusing on design strategies, communication and interaction modalities, nuanced activities, self-presentation, harassment, privacy, and self-disclosure. These recent works suggest that many questions still remain in social VR scholarship regarding how to ethically conduct research on these sites and which research areas require additional attention. Therefore, in this paper, we provide an overview of modern Social VR, critically review current scholarship in the area, raise ethical considerations for conducting research on these sites, and highlight unexplored areas. 
}

\maketitle

%% \section{Introduction} %for journal use above \firstsection{..} instead

\section{Introduction}
\textit{Social Virtual Reality} (VR) provides %is a 
novel digital spaces where users can interact, socialize, and game with one another through head-mounted displays (HMDs) \cite{mcveigh2018s,mcveigh2019shaping}. In these open-ended 3D virtual spaces, %worlds,
users engage in cultivating online social relationships\cite{zamanifard2019togetherness}, exploring diverse virtual activities %places 
\cite{maloneyActivities}, experimenting with self-representation \cite{freeman2020my,Freeman2020SelfRepresentation}, and enjoying immersive gaming\cite{maloneyChildren,maloney2020complicated}. These spaces %worlds 
are similar to traditional virtual worlds such as Second Life, Runescape, and Club Penguin. However, social VR provides a much more immersive experience due to the use of HMDs and open up new opportunities for embodied interaction. %Recent worldly events (e.g. the COVID-19 global pandemic) have demonstrated the increasing potential of using social VR for interaction and virtual relationships.
%use of social VR as a popular socio-technical phenomenon and emerging research agenda. 
The popularity and immense potential for social VR has %even 
signaled significant interest and investment from top technology companies such as Microsoft, HTC, and Facebook, who each own a %their own 
social VR platform. Recent worldly events (e.g., the COVID-19 global pandemic) have further demonstrated the increasing potential of using social VR for interaction and virtual relationships. As these platforms continue to grow in popularity, many questions still remain about the ramifications of social VR as it grows as an emerging research space. %especially since scholarship in this area is still in it's infant stages 

%social vr as an emerging research space, we are aware of the foll, to our knowledge there have been x papers published in the last two years
%\cite{jonas2019towards,mcveigh2018s,mcveigh2019shaping,sra2018your,moustafa2018longitudinal,maloney2020complicated,maloney2020nonverbal, maloneyActivities, maloneyChildren,zamanifard2019togetherness,maloneyAnonym,freeman2020my,Freeman2020SelfRepresentation,blackwell2019harassment,baker2019interrogating,tanenbaum2020make}.

Therefore, in this paper we provide an overview of modern %introduce 
Social VR, % currently the most widespread use of VR, provide an overview of 
critically review current scholarship in the area, %raise ethical considerations for conducting research on these sites, and highlight unexplored areas. 
%introduce \textit{Social VR}, an emerging and popular immersive socio-technical space, provide an overview of current scholarship in the area,
point researchers towards unexplored areas, and raise ethical considerations for how to ethically conduct research on these sites and which research areas require additional attention.

\section{\textit{Social} Virtual Reality}

\subsection{Defining and Characterizing Commercial Social VR}
For the purpose of this paper, we adopt and expand McVeigh et al.'s\cite{mcveigh2018s,mcveigh2019shaping} work to define %original 
social VR as any \textit{commercial} 3D virtual environment where multiple users can interact with one another through VR HMDs. We focus on commercial applications of Social VR for two %a few 
reasons. First, VR's relevance and societal impact will likely have a more significant impact on commercial VR uses than non-commercial uses.
%First, %one reason is that 
%the relevance and societal impact of VR will likely have a greater impact on %in 
%commercial VR uses rather than non-commercial uses. 
Second, %Another reason, is 
non-commercial platforms impose significant barriers to entry, including %largely 
programming expertise and VR knowledge that is not available to the average consumer. %and 
The majority of prior scholarship involving VR will also be more applicable to commercial social VR platforms (i.e., for 
entertainment purposes) as opposed to other forms of VR (e.g., for training and simulation purposes). %for VR). 
For example, commercial social VR is used in ways that researchers would not normally expect, such as users \textit{sleeping} in VR \cite{maloneyActivities}. This demonstrates that certain facets of social VR may be challenging to replicate in a laboratory.

Modern commercial social VR platforms afford four main characteristics,  %these characteristics 
making them attractive and popular digital social spaces. %socio-technical spaces. 
One of the characteristics is %are 
embodied virtual avatars. %, t
These avatars support full/partial body representation along with real-time movements and gestures.
%First, the use of embodied avatars, which support full/partial-body movements and gestures in real-time. 
 In these environments, users can create and customize their embodied avatars %which they use 
to explore different virtual spaces and interact %interacting 
with other users. The avatar body in social VR is the primary interface for the user.
Second, social VR supports vivid spatial and temporal experiences through high-fidelity 360-degree content and six degrees of freedom. 
These experiences include a plethora of activities and engagement experiences that may or may not resemble activities in the offline world. 
%For example, some activities such as meditation and yoga are offline activities done in social VR, but other activities such as RecRoyal are entirely
Third, there are multiple means of communicating and interpreting communication in social VR via verbal or non-verbal communication. The ability to communicate with another person via non-verbal means such as gestures and body movements %s 
has the potential to eliminate language barriers and allow for accessible communication to those who are impaired \cite{maloney2020nonverbal}. Finally, modern social VR platforms are free to play and are widely accessible to anyone who has a VR device; some social VR platforms are also accessible to non-VR users, allowing interaction between desktop-based users and VR users. To the best of our knowledge, most if not all, major social VR platforms are free to use. %, t
These include: AltspaceVR, RecRoom, VRchat, Facebook Spaces (discontinued), Facebook Horizon, Mozilla Hubs, Big Screen, High Fidelity, Sansar, Neos, Anyland, and Pokerstars.

\subsection{Popular Social VR Platforms}

The majority of social VR platforms afford various types of events/activities and the creation of diverse places and spaces.
%The majority of social VR platforms have events, places and space to create, and attend events. 
Each platform has different levels of customizable avatars and offers different levels of avatar fidelity. We introduce several of the most popular social VR platforms below. %some offering different levels of avatar fidelity. Below are a few popular platforms.
 
\textit{AltspaceVR.} Owned by Microsoft, AltspaceVR is described as ``the premier place to discover the next frontier of entertainment and community." In AltspaceVR, users can attend live events and meetups such as open mic night, improv comedy, meditation, yoga, LGBTQ meetups, or VR church. Users can also host their own events. Events on AltspaceVR can also be accessible via a PC.

\textit{RecRoom.} RecRoom is considered most popular among minors, while 
adults tend to prefer AltspaceVR \cite{maloney2020complicated}. The main activities in RecRoom are centered around games (e.g., paintball and basketball). Users can create their own private rooms. They can also venture into a central hub called the Rec Center and go into different rooms for gaming. 

\textit{VRchat.} Among all the social VR platforms, VRchat (owned by HTC) %among the others, 
affords minimal activities but features uniquely designed rooms (e.g., spaceship, Japan Shrine) that attract various types of users. %and the avatars has much more customization. 
It also offers the most sophisticated avatar customization compared to RecRoom and AltspaceVR. VRchat is ranked as one of the most popular applications on the Steam game marketplace.

\textit{Facebook Horizon.}
Currently, Facebook Horizon's is the newest of all the social VR platforms. %the above platforms, 
Horizon is Facebook's second social VR platform, building on the now discontinued Facebook Spaces. According to Facebook, Horizon is %described as 
a sandbox universe where users can create and craft their environments and games; they can also socialize with current Facebook friends or other users on the platform.

\subsection{Comparing Social VR with %Comparisons to 
Traditional Virtual Worlds}

Grounded in the definitions and characteristics of emerging social VR platforms, we therefore, summarize and highlight the similarities and uniqueness of social VR compared to traditional virtual worlds. 

The current designs and experiences of social VR platforms mimic behaviors and designs that were first observed in collaborative virtual environments (CVEs) \cite{benford1995user}. In fact, social VR can be compared to CVE in three ways: %has three main characteristics of CVE's: 
 multiple methods of communication, various interactive and collaborative experiences, and experiences mediated by the agency of digital self (i.e., avatar). 

\textbf{Communication.} In early CVEs, interactivity was mediated via rich text-based interaction \cite{bartle2004designing,bruckman1994programming}, involving role-playing, adventure, questing, and reenacting pop culture \cite{sanchez2009social, hoon2002lifestyle}. Collaboration was also facilitated in these spaces as they were seen as a more efficient means of communication in the workplace \cite{churchill1999virtual}. Similar to social VR, this facet was attractive to users because even though communication was mediated via text, it was \textit{real-time, unobtrusive, multi-user, and exclusive} \cite{evard1993collaborative}. However, two key differences should be noted between early text-based CVEs and social VR. First, %one, 
current interactions in social VR platforms are not yet \textit{archivable} as these in early CVEs were \cite{evard1993collaborative}. Second, %and two, 
text-based CVE systems limit the ability to convey and interpret rich social cues via verbal (e.g., voice intonation) or nonverbal behavior (e.g., gestures, facial expressions), while social VR significantly promotes this ability.

\textbf{Interaction and Collaboration.} The nature of \textit{sociality} in social VR very closely mimics early CVEs %as spaces where 
where users can be connected, play, and work with one another %, play with one another, and work with one another 
while being physically apart \cite{churchill2012collaborative}. As time progressed, CVEs evolved to include multiple means of communication specifically highlighting in-depth interpersonal communication \cite{bailenson2005digital,konijn2008mediated,bailenson2003interpersonal,bailenson2004transformed}.  These early examples demonstrate the nuances of CVEs when compared to face-to-face interaction and modern commercial social VR, yet the majority of prior scholarship was conducted in a controlled lab setting versus an open collaborative environment like social VR. 

As CVEs evolved to have a multi-player focus, different social groups began to emerge. For example, groups in MMORPGs vary in size but often have a common goal, enabling different types of social interactivity between group members \cite{mcewan2012m,nardi2006strangers}. %For example,
Some groups %also 
(often known as guilds) are highly organized, which promotes more dynamic and intimate relationships between guild members \cite{ducheneaut2006alone, ducheneaut2007life,nardi2006strangers}. Similar to social VR, the intimate relationships formed in these virtual groups can lead to social activities unrelated to actual in-game objectives, such as building substantial emotional bonds of friendship, intimacy, affection, and romance \cite{pace2010rogue,freeman2015simulating,freeman2016intimate}. 

\textbf{Avatar-Mediated Experiences}. Currently, facets like appearance and identity in most CVEs, MMORPGs, and virtual worlds are mediated via their embedded %an avatar 
systems. As avatars %these systems 
become more advanced and embodied, % hyper-personal 
 they may lead to %can create 
more nuanced connections with the users who control them. For example, customizing one's avatar is commonly considered an activity in virtual worlds \cite{huh2010dude, schroeder2012social}. This activity is also seen in social VR \cite{freeman2020my, Freeman2020SelfRepresentation}. In particular, social VR creates a more natural avatar-body connection, where the users' body acts as the interface rather than a mouse or keyboard. This connection may lead to more naturalistic experimentation with different bodies as compared to experimenting in virtual worlds \cite{ducheneaut2009body,freeman2015simulating,freeman2016revisiting}.
%Although avatar mediation is generally facilitated via mouse or controller, this still led to connections of experimentation with a different digital body \cite{ducheneaut2009body,freeman2015simulating,freeman2016revisiting}. 
 
In summary, social VR demonstrates both similarities and uniqueness compared to traditional CVEs. This leads to an emerging research space on social VR about how these novel digital systems are shaping our online social lives. %prior scholarship in CVEs demonstrate both distinct similarities to modern social VR, yet many questions still remain about the effects and potential differences between social VR and CVEs.

\section{Current Social VR Research Landscape}

There is a growing research agenda to explore social VR as an emerging technology. 
%Scholarship in social VR is still in the infant stages but c
Current studies have focused on design strategies \cite{jonas2019towards, mcveigh2019shaping, sra2018your}, communication and interaction modality \cite{mcveigh2018s, moustafa2018longitudinal, maloney2020nonverbal, tanenbaum2020make, maloneyActivities, baker2019interrogating}, long-distance couples' and children's experiences \cite{zamanifard2019togetherness, maloney2020complicated, maloneyChildren}, exploration of self-representation \cite{freeman2020my,Freeman2020SelfRepresentation}, harassment \cite{blackwell2019harassment}, and privacy \cite{maloneyAnonym}.

\subsection{Design Strategies}
Early scholarship surrounding design strategies for social VR was conducted by McVeigh et al.\cite{mcveigh2018s,mcveigh2019shaping}. In their works, %this work that 
McVeigh et al. coined the term \textit{social VR} and introduced the design strategies surrounding this pro-social interaction in social VR. In particular, McVeigh et al. highlighted three types of recommendations: (1) leveraging offline modalities and interactivity, (2) self-governed social environments and safe onboarding, and (3) creating meaningful connections with friends, inside and outside of the environment. However, as these recommendations were not exhaustive, they did not point towards future design directions for accessibility, mitigating unwanted interactions, and designing for different demographics (e.g., age). 

Similarly, design strategies by Jones et al. \cite{jonas2019towards} only focused on the taxonomy of social VR design, which did not include solely commercial social VR applications but rather a broader definition of social VR %defined 
as ``a growing set of multi-user applications that enable interactivity between head-mounted displays." In this work, Jones at al. Shed light on three core features of 
social VR: avatars, interaction with others, and the environment. However, these facets did not fully explain  why the immersiveness of social VR made these facets any different from prior work done in traditional virtual worlds or comment on how they may influence interactivity in social VR.

%mcveigh
%jonas
%sra

\subsection{Communication and Interaction Modalities}
One of the nuances of social VR is the interpretation of interpersonal communication cues, which have been demonstrated to mimic behavioral cues of the offline world\cite{moustafa2018longitudinal,mcveigh2018s,maloney2020nonverbal,tanenbaum2020make}. McVeigh and colleagues' early work highlighted that the affordances of locomotion, spatial embodiment, and social mechanics were key modalities for expressing \textit{sociality} on social VR platforms \cite{mcveigh2018s}. Similar work by Moustafa and Steed demonstrated that users perceived interactions in social VR as identical to those in the offline world, specifically relating to gestures and embodiment\cite{moustafa2018longitudinal}. 
These studies demonstrated the distinct similarities of interpersonal communication between social VR and the offline world. Still, they did not fully explain how or why users were using communication modalities in social VR.In addition, Maloney et al. demonstrated that social VR users used non-verbal communication modalities to communicate in a safer and more comfortable fashion  \cite{maloney2020nonverbal}. This work also highlighted some key interaction outcomes of using non-verbal communication in social VR. Although 
this work sheds light on how a particular subset of social VR users (e.g., %i.e., 
marginalized users) may take advantage of non-verbal communication, it did not classify in-depth all the different types of non-verbal communication modalities such as facial control mentioned %classified 
by Tanenbaum and colleagues \cite{tanenbaum2020make}.

%mcveigh, moustafa2018
%baker2019

%\cite{mcveigh2018s, moustafa2018longitudinal, maloney2020nonverbal, maloneyActivities, baker2019interrogating}
\subsection{Nuanced Activities \& Engagement}
Social VR inherently affords a broad range of social activities and engaging %engagement 
experiences. Many are traditional activities that can be found in virtual worlds and MMORPGs, such as game playing, entertainment, and learning \cite{maloneyActivities,maloneyChildren}. However, one uniqueness of engaging in social VR is re-experience %one commonality in prior scholarship is re-experiencing 
offline actions, events, and interactivity. %but in social VR. 
For example, Zamanifard and Freeman highlighted that long distance couples used social VR as a means to be %of \textit{being together} and feeling 
connected and replicate their offline activities so as to feel virtually %and using the affordance of the body to feel 
\textit{together} \cite{zamanifard2019togetherness}. %Scholarship by 
Maloney et al. also highlighted that mundane offline activities such as \textit{dancing} and \textit{sleeping} were %which are 
re-experienced in social VR but in a new way %such as \textit{dancing} and \textit{sleeping} 
\cite{maloneyActivities}. Similar to Zamanifard and Freeman's work, the affordance of the body as the sole interface made behaviors such as dancing more engaging in social VR and even as a means to invite interactivity. Additionally, behaviors such as sleeping demonstrate the distinct affordances of social VR over traditional virtual worlds and online experiences. In the same study, Maloney et al. found that users' enjoyed a variety of activities including social and mental improvement, immersive cultural appreciation, and engaging in immersive events. More scholarship is needed to confirm these findings and extrapolate different immersive activities.

Another set of activities in social VR are geared towards  younger users. %demographic. 
For example, Maloney et al. described social VR as a new experience for relationship building between different generations \cite{maloney2020complicated} and a new modality of building intimacy and stronger emotional connections \cite{maloneyChildren}. These works %This work 
also highlighted the inter-cultural exchange between younger users and adult users as an everyday activity. 
Therefore, activities in social VR have emerged beyond gameplay and entertainment and expand to %but 
education, relationship building, and immersive specific behaviors (e.g., dancing and sleeping).

%long distance couples
%childrens experiences
%chiplay activities paper

\subsection{Self-Presentation \& Avatar} %Perception}
Another focus in social VR scholarship revolves around %area of early exploration in social VR scholarship surrounds 
how users choose to present themselves in social VR. Like
traditional MMORPGs and virtual worlds, social VR users opt for a variety of choices in avatar self-representation. Yet recent works by Freeman et al. \cite{Freeman2020SelfRepresentation,freeman2020my} highlight that users choose to represent themselves similarly to their physical selves %real identity 
or craft a self-presentation %representation 
based on the affordances and social atmosphere of the specific platform. Therefore, these works explore a different aspect of selective self-presentation that emphasizes consistency and involves an interplay of body, avatar, audience, and conscious personal choice. They have also highlighted how such nuanced self-presentation in social VR affects users' understandings of self. For example,
%The differences between the two lie in the particular experiences of users for example, marginalized members choosing to create an avatar different from their offline identity. Additionally, 
social VR avatars allow for a more in-depth exploration of identities, particularly by trans users. For them, this was an embodied approach of exploring their non-traditional gender identity.

\subsection{Harassment, Privacy, and Self-Disclosure}

%vrst paper
%blackwell
%childrens paper

%introduce social VR talk in depth about what it is and popular platforms
Similar to traditional virtual worlds, negative experiences such as harassment and unwanted interactivity occur in social VR. %In particular, 
For example, a survey by Shriram and Schwartz highlighted that 
two out of seven women and 21 out of 99 men reported that they had experienced harassment in social VR \cite{shriram2017all}. %Later w
Work by Blackwell et al. pointed out that feelings of presence, body movement, embodiment, and voice chat could aggravate harassment \cite{blackwell2019harassment}. %, h
However, it should be noted that this study only explored harassment in Facebook Spaces. More research on other social VR spaces is needed to verify such findings. % and scholarship is need in other social VR spaces. 
Additionally, Maloney et al. demonstrated that marginalized users (e.g., younger users, non-English speakers, %non-english speakers, 
LGBTQ, and women) are often subject to more harassment based on their gender, %identity, 
sexuality, race, and age, with one instance of virtual sexual assualt towards a minor \cite{maloney2020nonverbal, maloneyChildren, maloney2020complicated}. 

Thus far, the area of privacy and self-disclosure in social VR has also revealed how different users perceive and approach privacy and self-disclosure in various ways 
%unique differences between how users perceive privacy and self-disclosure 
\cite{maloneyAnonym}. In particular, Maloney et al. described three patterns regarding self-disclosure in social VR: (1) sharing information based on familiarity with friends or close acquaintances; (2) preferring to remain anonymous when sharing information; and (3) open to sharing information regardless of their familiarity and anonymity. These findings demonstrate particular trade-offs and conflicting points of view %points 
regarding privacy and self-disclosure in social VR. Future directions for privacy consent and privacy transparency are needed as the use of social VR %use 
continues to grow.

%distinct tradeoffs for users. For example, Maloney et al. demonstrate 

\section{Ethical Considerations from Prior VR and Virtual Worlds Research} %Research Concerns \& Risks of Social VR}
Existing social VR literature %The above mentioned literature demonstrates 
has highlighted future areas of exploration in this emerging research space. %for research. Yet, 
However, we express cautions and urge %heed 
researchers to draw ethical considerations from both prior VR work \cite{slater2020ethics,madary2016real, slater2014grand,behr2005some} and work in traditional virtual worlds as they conduct future social VR research \cite{mckee2009playing,grimes2009virtual,boellstorff2012ethnography}. %This is two-part, one in as 
These considerations are two-folded; one focuses on VR and virtual worlds as a research context and the other focuses on the safety of VR and virtual worlds % a general safety context 
for consumer use. 

\subsection{Challenges for Conducting VR Research} %Research Considerations in Prior VR Scholarship}
%Research considerations for VR have been of growing concern as the technology has 
As VR technology has become widely available and accessible, challenges for conducting VR research are of growing concern. Slater introduced %these concepts as 
two grand challenges for VR technology \cite{slater2014grand}, which focus on consumer well-being. The first challenge recognizes %, recognizing 
that VR will become a mass consumer product. Therefore, %and such he outlined that 
these devices must be cheap, safe, and deliver compelling experiences so that %and 
researchers should investigate their longitudinal impacts on consumers. Second, barriers and challenges determined by the offline-world physics %which 
may impact user experiences. %These challenges and ethical concerns lie within the uses of consumer well-being. 

Regarding the safety and experimentation of users, Behr and colleagues insisted that \textit{motion sickness} must be mitigated and that researchers must assist their research participants to reorient to the offline world \cite{behr2005some}. To address the posed risks of motion sickness, Behr et al. suggested that exposure time should be limited until adaption to VR has occurred; tasks prone to sickness should be avoided, and considerations for VR use should be on an individual basis. Yet, these suggestions did not fully elaborate on other ethical concerns or other risks posed by VR. Madary and Metzinger built upon Behr et al.'s work and highlighted six potential issues of VR research: (1) %\textit{
limits of experimental environments, (2) informed consent with regard to the lasting psychological effects, (3) risks associated with clinical applications of VR, (4) the possible use of VR research for malicious purposes, (5) online research using VR, and (6) inherent limitations of a code of conduct for VR research \cite{madary2016real}. Madary and Metzinger also highlighted that ethical VR experimentation must follow the principles of \textit{beneficence} and \textit{non-maleficence} in VR. To address the concerns of VR and concerns regarding VR research, Madary and Metzinger proposed %a 
the following recommendations. First, all research must follow procedures of informed consent and preserve participants' autonomy and trust. Second, be honest and clear %clarity 
with the scientific progress of VR for medical treatment.

\subsection{%Notable 
Concerns \& Risks Towards Consumers in VR Research} % Scholarship}
Prior VR scholarship has highlighted several risks and concerns towards consumers %A few risks towards consumers have emerged in prior VR scholarship 
\cite{slater2020ethics, madary2016real,adams2018ethics,maloney2019ethical}, which we summarize below.  %, b
%Below we briefly summarize the risks and concerns of previous scholarship. 

One risk is %are 
the \textit{cognitive, emotional, and behavioral changes via virtual embodiment}. %, f
For example, embodied virtual avatars are capable of altering a person's racial biases \cite{maister2015changing,Banakou2013,banakou2016virtual,peck2013putting,peck2018effect,maister2013experiencing}. Another risk is the \textit{over-use of VR content}, especially regarding the frequency and duration of using the device. Additionally, \textit{negative psychological effects} may emerge when users leave VR (e.g., fantasy-based content).%of leaving (e.g., fantasy) VR}, u
Users may feel less enthusiastic about venturing back into the offline world and suffer depression or withdrawals. As negative experiences may occur in VR, concerns surrounding the \textit{legal and ethical responsibilities} of VR also begin to emerge as this is largely a grey area. %grey-area. 

The area of \textit{privacy} is also of growing concern of VR use, specifically regarding the sharing of personal data with third parties. Similar to other online technologies (e.g., social media), users will have access to %\textit{
negative content and/or mature content that is % which will be 
readily available.%, t
This is of concern as immature audiences such as children and teens may have access to this content. %Additionally, with the above mentioned concerns 
Slater et al. even highlighted that some if not all of these concerns may become more apparent as VR moves towards higher levels of realism \cite{slater2020ethics}.

To address the above-mentioned risks, Slater and colleagues proposed five principles for action, including: (1) Minimizing potential harm of immoderate use, (2) minimizing content-induced risk, (3) selecting levels of deception, (4) educating implementers and participants, and (5) protecting personal information. 

\subsection{Challenges for Conducting Virtual Worlds Research} %Research Considerations in Prior Virtual Worlds Scholarship}
The research potential of virtual worlds have been a long-standing interest  of scientific communities \cite{bainbridge2007scientific,marsh2010young,di2008can,kafai2017designing,kolodner2017drawn,bruckman1997moose,sanchez2009social, hoon2002lifestyle,bartle2004designing,bailenson2005digital,nardi2006strangers,ducheneaut2006alone, ducheneaut2007life, hendaoui20083d, minocha2010conducting,mckee2009playing, grimes2009virtual}. Bainbridge explains that virtual worlds have emerged as an interdisciplinary space \cite{bainbridge2007scientific}.%, h
He also highlighted %raised 
emerging challenges for conducting %regarding 
research in these spaces. One challenge lies in % area was 
the psychological concerns about %of 
the attachment with %towards 
a users' virtual avatar/character. Another challenge focuses on % area  of concern was 
ethics regarding human subjects research in virtual worlds. %and when researchers are subject to institutional human subjects review boards.

Another set of considerations by Minocha et al. highlighted their experiences and suggestions for conducting empirical research in Second Life \cite{minocha2010conducting}. One of the research considerations mentioned is %how Minocha et al. suggest 
communicating details of the experiment with said ethics committee and%, specifically, 
focusing on how to achieve %
the \textit{privacy and dignity of observations} in public and private spaces in virtual worlds. Another consideration emphasized % was 
the recruitment of participants, where they scouted potential participants within Second Life and sent them a short, direct message informing them of the potential research opportunity. Minocha et al. also suggested that researchers should be part of the online community for an extended period of time before any formal data could be %is 
collected. The goal is to %so as to 
familiarize themselves with the community before conducting any research. %Another key consideration was that 
Researchers were also recommended to develop their identity in the virtual worlds % researcher's identity on the platform, 
in order to maintain participants' trust and confidence in the project. Finally, Minocha et al. stated that researchers should adhere to the platform-specific community standards of practice.  

Grimes et al. also highlighted a few considerations when thinking of approaching research in virtual worlds \cite{grimes2009virtual}. One consideration is that research on virtual worlds cannot be conducted exactly as research in offline space. Another consideration is that respecting individuals' privacy and their avatars are essential to build and keep participants' trust.  Finally, striving for transparency and respecting the (social) norms of the virtual community should be essential to any virtual world research. 

In addition, %\textbf{Notable Concerns \& Risks Towards Consumers in Prior VR Scholarship.} A few 
several ethical and legal considerations regarding concerns and risks towards consumers in virtual worlds research have also been observed %of consumer use has been observed in traditional virtual worlds 
\cite{hendaoui20083d}. One example is the intellectual property rights of user-generated content, %creations 
such as objects created by users. % or the varying rules on different platforms. 
Additional concerns emerged regarding how offline laws apply to virtual worlds. For example, regarding privacy and safety, at what point does misrepresentation (via avatar) become unethical or criminal? %Another concern is w
What are the ramifications of the incorporation of copyrighted music/video that occur in virtual worlds?

%although we have social vr research the ethical concerns in this space is understudied. no prior studies have talked about ethics.
%ethical concerns are a longstanding issue for vr and virtual worlds.
%social vr brings in new challenges and new concerns. 
%VR:
%https://www.frontiersin.org/articles/10.3389/frvir.20200001/full?
%Virtual Worlds:
%McKee, H. A., & Porter, J. E. (2009). Playing a good game: Ethical issues in researching MMOGs and virtual worlds. International Journal of Internet Research Ethics, 2 (1), 5-37.
%Grimes, J. M., Fleischman, K. R., & Jaeger, P. T. (2009). Virtual guinea pigs: Ethical implications of human subjects research in virtual worlds. International Journal of Internet Research Ethics, 2(1), 38-56.
%Boellstorff, T., Nardi, B., Pearce, C., & Taylor, T. L. (2012). Ethnography and virtual worlds: A handbook of method. Princeton University Press.

%1. introduction
%2. social vr  — what is social VR, what are the main platforms, what are the unique affordance compared to traditional virtual worlds
%3. current social Vr research landscape
%4. emerging ethical research concerns in social Vr — ethical concerns in prior vr and traditional virtual worlds; ethical concerns in social VR due to its uniqueness and what have emerged in prior work on social VR

\section{Proposed Guidelines for Conducting Research in Social VR}
Grounded in %on 
prior scholarship on challenges, concerns, and risks toward %ethical considerations for 
VR and virtual world research and taking the nuances of social VR technology into account, %in of VR and virtual worlds, 
we propose the following guidelines %considerations 
for conducting %ethical research in 
social VR research in an ethical way. %It should be noted that 
%%These guidelines %considerations are neither complete nor exhaustive but aim to lead to open conversions and reflections. %as they are considerations emerging from prior scholarship. Our suggested guidelines are:

\begin{itemize}
  \item \textbf{The welfare and consent of the research participant(s) should be prioritized in all social VR research}. %\cite{world2001world,national1978belmont}. 
  
  Specifically, social VR research should adhere to \textit{beneficence} and \textit{respect for persons} \cite{world2001world,national1978belmont}. This requires %Thus, 
  maximizing potential benefits and minimizing possible harms and making sure participants have autonomy and rights to make their own decisions. Informed consent of risk must be given and only should be waived if complying with platforms' policies. 
  %Additionally, anonymity and privacy of participants' must be kept safe and not reveal their identity (i.e. avatar). 
  
  \item \textbf{Knowledge of said Social VR platform}
  
  Before embarking on a specific social VR to conduct research, %platform, 
  researchers should aim to spend ample time on the platform to fully understands the nuances, novelty, and uniqueness of the platform. As each platform has different norms, user-bases, and affordances, such knowledge is necessary to ensure that the researcher understand the underlying culture and social atmosphere of the specific platform.
  
  \item \textbf{Ensuring Privacy and Care of Participants}
  
  Participants' anonymity and privacy must be kept safe as not to reveal their identity unless given consent by participants. This includes data such as gait, motion-tracked data, avatar appearance, username, and voice. For example, as scholarship demonstrates, motion-tracked data can be identifiable up to 95\% accuracy when other personally identifiable information is redacted \cite{miller2020personal}. This demonstrates that researchers should  take measures to ensure the privacy of unique motion signatures (e.g., biometric data, eye-tracking, brain-computer interface) of participants.  An additional level of care should be noted when conducting research with users and particularly younger users on these platforms, as users are not always entirely aware of said risks posed. For example, when conducting research with adolescents seeking informed consent of the younger user and their parent/guardian and only waiving consent when minimal to no harm is possible.

  %How does this unique motion signature coupled with other biometric data (e.g., eye-tracking, brain-computer interfaces) ensure participants' privacy and safety.
  
  \item \textbf{Compliance with Platforms' Terms of Service}.
  
  Researchers must aim to comply with the specific social VR platform's terms of service when conducting research as this may affect their research methodology. For example, if a researcher aims to conceal or present a different identity in a specific community, it may not comply with certain platforms' community rules. For example, AltspaceVR's terms of service %in AltspaceVR 
  restricts creating ``a false identity or impersonating another person or entity in any way." Similarly, RecRoom also restricts ``impersonate or misrepresent your affiliation with any person or entity." These platforms' terms of service also include procedures for conducting research. For example, public spaces on some platform are fair ground for conducting research, whereas private rooms are not.

  %\item \textbf{Knowledge of said Social VR platform}
  
  %Before embarking on a specific social VR research platform, researchers should aim to spend ample time on the platform as each platform has different norms, user-bases, and affordances. 
  %why would you need to know the platform in order to conduct ethical research. for example Recroom more geared towards childrenen

  \item \textbf{Respect for Community Norms}
  
  Most social VR platforms foster %have 
  specific sub-communities (e.g., users of a certain identity and/or interest). Some sub communities are formally recognized by said platforms whereas others are created and managed by their respective group members. Sub-communities may %each which 
  have their own rules, social norms, and regulations. Researchers should respect these communities and adhere to their community norms while conducting research. Some sub-communities may also be cautious when %groups as some groups do not appreciate 
  researchers conduct research covertly within their communities. 
  %ethnographic approach as an alternative where the norms are considered. 

  \item \textbf{Recruit diverse participants}
  
  Social VR attracts a broad range of users, each with different demographics relating to race, gender, sexual orientation, ability, and age. Researchers should seek to diversify participants as to best ensure that diverse viewpoints and perspectives are taken into account. The current lack of representation was demonstrated by Peck et al. who showed that female participants are significantly underrepresented in VR research \cite{peck2020mind}.
  %Women's representation in Vr research would be great to include Peck paper

\end{itemize}

%The above-mentioned guidelines are grounded in recommendations from prior VR and virtual worlds scholarship. It should be noted that 

We acknowledge that these guidelines %considerations 
are neither complete nor exhaustive but aim to lead to open conversions and reflections. We also note that the uniqueness of social VR may continue to lead to new concerns and emerging guidelines that are currently not included in this paper.  % prior scholarship. 
For example, as advances in VR technology further enable photo-realistic telepresence, what considerations should be made for participants as social experiences in VR become indistinguishable from the offline world? These concerns, therefore,
%These set of guidelines and future research considerations point towards 
point to unexplored areas and emerging questions for social VR research, which we discuss in the next section.

%\textbf{Add here: highlight how the uniqueness of social VR may lead to new concerns/guidelines that are not included in prior work in VR/virtual worlds}

%\textbf{Transition here from this research, why do guidelines lead to unexplored areas, questions etc}

\section{Call to Action for Social VR Researchers}

Our guidelines for conducting social VR research in an ethical way also highlights the need to further %Conducting ethical research is one half of the scientific process; the other half lies in investigating 
investigate unexplored or understudied areas in social VR research. %or briefly investigated areas. 
In this section, we point researchers towards such areas. %of social VR that have little exploration and leave more to be explored. 
These emerging questions are grounded in research considerations in prior social VR scholarship surrounding %which surrounds 
social VR design, self-presentation, safety, well-being, immersive experiences, and privacy, 

%The design of social VR systems have been explored by researchers \cite{jonas2019towards, mcveigh2019shaping, sra2018your}, but require more investigation, specifically:

\begin{itemize}
  \item What design considerations should be made to help platforms maintain intimate communication and still grow their user base?
  \item What does the design of social VR platforms lend itself to? Should social VR mimic the offline world or be a balance of both the offline and online world?
  \item What are the considerations for designing a social environment %designed 
  inclusively for marginalized groups of all kinds?
  \item How can bias (e.g., gender/racial), body dysmorphia, and other concerns of self-identity be mitigated in social VR?
  \item How do we help protect personal space and other considerations of psychological and physical vulnerability in social VR?
  \item How can you validate the identity of another person in social VR without violating their privacy?
  \item Given the immense sociality of social VR, what are the long-term psychological and behavioral effects of social VR immersion? 
  \item What are the legal considerations and ramifications regarding content, interactivity, misbehavior, and privacy in social VR?
  
  \item What information is considered biometric data in social VR, and what types of information is not biometric but can still be identifiable?
\end{itemize}

\section{Conclusions}
In this paper, we call attention to commercial \textit{social VR}, a popular online digital space where users socialize via HMDs in immersive virtual worlds. These spaces resemble traditional 2D virtual worlds where users engage in interactive activities and are represented via avatars. Yet, social VR affords facets that traditional 3D %2D 
virtual experiences do not offer, such as full/partial body tracked avatars and 360-degree immersive content. %and identity experiences.
Despite the increasing popularity of social VR, %Although a popular technological phenomenon, social VR 
research in this area is still emerging. Therefore, in this paper, we provide a summary of social VR scholarship relating to design strategies, communication and interaction modalities, nuanced activities and self-presentation, harassment, privacy, and self-disclosure. Additionally, we provide ethical guidelines on how to conduct research in this space and point the VR community towards unexplored areas of social VR. %and  
We hope that our summary, guidelines, and research directions motivate %researchers towards investigating ways in which will 
and inform future directions for designing more safer, transparent, and fulfilling social VR experiences.

\bibliographystyle{abbrv-doi}

\bibliography{template}
\end{document}